\documentclass[12pt]{article}
\usepackage{fullpage}
\usepackage[]{graphicx}

\begin{document}

\title{Cooling many particles at once}
\author{Almut Beige,$^{1,2}$ Peter L. Knight,$^2$ and Giuseppe Vitiello$^3$ \\[0.2cm]
{\small \em $^1$Department of Applied Mathematics and Theoretical Physics,} \\[-0.2cm]
{\small \em University of Cambridge, Wilberforce Road, Cambridge CB3 0WA, UK} \\[-0.2cm]
{\small \em $^2$Blackett Laboratory, Imperial College, Prince Consort Road, London SW7 2BW, UK} \\[-0.2cm]
{\small \em $^3$Dipartimento di Fisica ``E. R. Caianiello,'' I.N.F.N. and I.N.F.M.,} \\[-0.2cm]
{\small \em Universit\'a di Salerno, 84100 Salerno, Italy}}

\date{\today}

\maketitle

\begin{abstract}
We propose a mechanism for the collective cooling of a
large number $N$ of trapped particles to very low temperatures
by applying red-detuned laser fields and coupling them to the
quantized field inside an optical resonator. The dynamics is
described by what appears to be rate equations, but where some
of the major quantities are coherences and not populations.
The cooperative behavior of the system provides cooling rates
of the same order of magnitude as the cavity decay rate $\kappa$.
This constitutes a significant speed-up compared to other
cooling mechanisms since $\kappa$ can, in principle, be as
large as $\sqrt{N}$ times the single-particle cavity or
laser coupling constant. \\[0.2cm]

Pacs numbers: 03.67.-a, 42.50.Lc
\end{abstract}

\maketitle

\section{Introduction}

Cooling and trapping techniques have improved dramatically over
the last few decades. A very efficient method to transfer, for
example, a single atom rapidly to a very low temperature is
sideband cooling \cite{diedrich}. This requires a red-detuned
laser field whose detuning equals the frequency of the vibrational
mode of the atom. When the laser excites the system from the
ground to an excited state,  the vibrational energy reduces by one
phonon. Afterwards, the atom most likely returns into the ground
state via spontaneous emission of a photon and without regaining
energy in the vibrational mode. The corresponding non-unitary
evolution effectively reduces the temperature of the atom and
yields an overall decrease of the von Neumann entropy in the
system \cite{Bartana}. Other experiments aim at cooling molecules,
which have a much richer inner level structure and are therefore
harder to control than atoms.

\begin{figure}
\begin{minipage}{\columnwidth}
\begin{center}
\resizebox{\columnwidth}{!}{\rotatebox{0}{\includegraphics{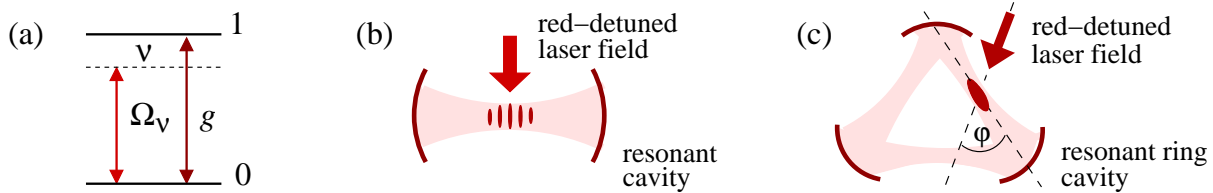}}}
\end{center}
\vspace*{-0.5cm} \caption{ Atomic level scheme (a) and
experimental setup for the collective cooling of many particles
trapped inside an optical cavity (b) or an optical ring resonator
(c).} \label{setup}
\end{minipage}
\end{figure}

Here we apply the idea of side band cooling to a large number of
particles (atoms, ions or molecules). As in the one-atom case, a
red-detuned laser field excites the particles, thereby
continuously reducing the number of phonons in the system. To
return the particles into their ground state, they should couple
to the quantized field of a leaky optical cavity, whose frequency
$\omega_{\rm cav}$ equals the dipole transition frequency
$\omega_0$ of each particle  (see Figure \ref{setup}(a)) . Once
the particles transferred their excitation into the resonator
mode, it leaks out through the cavity mirrors. Using cavity decay,
instead of spontaneous emission from excited levels, helps to
avoid heating due to rescattering of photons within the sample. It
also minimises spontaneous emission into unwanted states and
allows to control even complicated level structures, like
molecules.

Crucial for obtaining maximum cooling is the generation of {\em
cooperative} behavior of the $N$ particles in the excitation step
as well as in the de-excitation step. This is possible when the
Rabi frequency $\Omega_\nu$ of the laser field for the cooling of
a vibrational mode with frequency $\nu$ is for all particles the
same and all particles see the same cavity coupling $g$ (small
variations of $g$ and $\Omega_\nu$ and non-ideal initial
conditions do not substantially affect our conclusions and will be
considered elsewhere \cite{new}). Realising this inside an optical
resonator requires self-organization of the particles in the
antinodes of the cavity field, as predicted in
Ref.~\cite{peter,nonsense}.  If the dipole moments of the atoms
are in the average parallel to the cavity mirrors, the Rabi
frequencies $\Omega_\nu$ are practically and to a good
approximation for all particles the same, if the laser field
enters the cavity from the side as shown in Figure \ref{setup}(b).
The laser could also enter the cavity through one of the cavity
mirrors, as it was the case in the many-atom cavity QED experiment
in Grangier's group in 1997 \cite{Grangier}.  Alternatively, a
ring resonator can be employed, as in
Refs.~\cite{peter,Nagorny,jaksch,Kruse}, if a laser field with
$\omega_{\rm laser}$ enters the resonator in a certain
angle\footnote{This assures that the phase factors in the cavity
interaction term of the Hamiltonian can be absorbed into the
definition of the excited states $|1 \rangle$ of the particles in
the respective positions. The same applies for the phase factors
of the laser amplitudes, which are now exactly the same as the
phase factors of the cavity field term.} $\varphi$ with $\cos
\varphi = \omega_{\rm laser}/\omega_{\rm cav}$ (see Figure
\ref{setup}(c)).

We also require that the particles are initially all prepared in
their ground state (in the large $N$ limit fluctuations can be
neglected). Then the time evolution of the system remains
restricted within a Dicke-symmetric subspace of collective states
\cite{Dicke}. As shown below, these states experience a very
strong coupling to the laser field as well as to the cavity mode
and the system evolves into a stationary state with no phonons on
the time scale given by the cavity photon life time. Several
schemes for the cooling of atomic ensembles have been proposed
\cite{vuletic0,vuletic1,horak,peter,jaksch} and first
cavity-cooling experiments have already been performed
\cite{vuletic,Nagorny,Kruse}. Compared to these, the scheme we
 propose  provides a significant speed-up of the cooling process
if operated in a regime with
\begin{eqnarray} \label{reg}
\kappa \, \sim \, \sqrt{N} g \, , ~ {\textstyle {1 \over 2}} \sqrt{N}
\eta \Omega \, \gg \, \Gamma ~~ {\rm with} ~~
\Omega \equiv \Big( \sum_\nu \Omega_\nu^2 \Big)^{1/2} \, .
\end{eqnarray}
Here $\kappa$ denotes the cavity photon decay rate, $\Gamma$ is
the spontaneous decay rate of a particle in the excited state and
$\eta$ is the Lamb-Dicke parameter characterizing the steepness of
the trap. For sufficiently large $N$, condition (\ref{reg}) can be
fulfilled even if $\eta \ll 1$ and the system is operated in the
bad cavity limit with $\kappa \gg g^2/\kappa > \Gamma$. Moreover,
we remark that Eq.~(\ref{reg}) describes a {\em strong damping
regime} in which the system can only accumulate a small amount of
population in the excited states of the particles. 

The proposed cooling scheme can be used to cool a large number of
particles very efficiently.  It is therefore an interesting
question, whether the described setup might be used for the
preparation of Bose-Einstein condensates.  Currently, these
experiments mainly use evaporative cooling \cite{evaporative}
which systematically removes those atoms with a relatively high
temperature from the trap. Consequently, only a small percentage
of the initially trapped atoms is finally included in the
condensate. If one could instead cool all the atoms efficiently,
yet at the same time avoid the loss of particles, it should become
easier to experiment with large condensates. Besides, cooling is
also crucial for ion trap quantum computing where the achievable
gate operation times can depend primarily on the efficiency of the
cooling of a common vibrational mode \cite{cold}.

\section{Bosonic behavior of a large atomic sample} \label{sec2}

We consider a collection of $N$ two-level particles with ground
states $|0 \rangle_i$ and excited states $|1 \rangle_i$, each of
them described by $\sigma_{3i} = {1 \over 2} ( |1 \rangle_{ii}
\langle 1| - |0 \rangle_{ii} \langle 0|)$ with eigenvalues $\pm {1
\over 2}$. Transitions between the two levels are generated by
$\sigma_i^+=  |1 \rangle_{ii} \langle 0|$ and $\sigma_i^-=  |0
\rangle_{ii} \langle 1|$. Our fermion-like $N$-body system is thus
described by the su(2) algebra
\begin{equation}
\label{su2} [\sigma_3, \sigma^\pm] = \pm \sigma^\pm \, , ~~~
[\sigma^-,\sigma^+] = - 2 \sigma_3
\end{equation}
with $\sigma^\pm = \sum_{i=1}^N \sigma_i^\pm$ and $\sigma_3 =
\sum_{i=1}^N \sigma_{3i}$. Under the action of $\sigma^\pm$,
describing the laser excitation, the initial state with all
particles in the ground state, $|0 \rangle_{\rm p}$, is driven
into the Dicke-symmetric states $|l \rangle_{\rm p}$ with
\begin{eqnarray} \label{L}
|l \rangle_{\rm p} &\equiv& \big[ \, |0_1 0_2 ... 0_{N-l} \, 1_{N-l+1}
1_{N-l+2} ... 1_N \rangle + . . . + |1_1 1_2 ... 1_l \,
0_{l+1} 0_{l+2} ... 0_N \rangle \, \big] / {\textstyle
\sqrt{N \choose l}} \, , ~~~
\end{eqnarray}
a superposition of all states with $l$ particles in $|1 \rangle$.
The difference between excited and unexcited particles is counted
by $\sigma_3$ since $_{\rm p} \langle l| \sigma_3 |l \rangle_{\rm
p} = l - {1 \over 2}N$. For any $l$
\begin{eqnarray} \label{rel}
\sigma^+ \, |l \rangle_{\rm p} &=& \sqrt{l+1} \, \sqrt{N-l} \,
|l+1 \rangle_{\rm p} \, , \nonumber \\
\sigma^-  \, |l \rangle_{\rm p} &=& \sqrt{N-(l-1)} \, \sqrt{l} \,
|l-1 \rangle_{\rm p} \, ,
\end{eqnarray}
showing that $\sigma^\pm$ and $\sigma_3$ are represented on $|l
\rangle_{\rm p}$ by the  Holstein-Primakoff non-linear boson
realization \cite{Holstein,SUV}
\begin{eqnarray} \label{sigma}
\sigma^+ = \sqrt{N} S^+ A_{S} \, , ~~ \sigma^- = \sqrt{N} A_{S} S^-
\, , ~~ \sigma_3 = S^+ S^- - {\textstyle {1\over 2}} N
\end{eqnarray}
with
\begin{eqnarray}
A_{S} = \sqrt{1 - S^+ S^-/N} \, , ~~ S^+  |l \rangle_{\rm p} =
\sqrt{l+1} \, |l+1 \rangle_{\rm p} \, , ~~
S^- |l \rangle_{\rm p} = \sqrt{l}  \, |l-1 \rangle_{\rm p}
\end{eqnarray}
for any $l$. The $\sigma$'s
still satisfy the su(2) algebra (\ref{su2}). However, for $N \gg
l$, Eqs.~(\ref{rel}) become
\begin{equation} \label{relc}
\sigma^\pm \, |l \rangle_{\rm p} =  \sqrt{N} \, S^\pm \, |l
\rangle_{\rm p}
\end{equation}
and thus $S^\pm = \sigma^\pm/\sqrt{N}$ for large $N$. In the large
$N$ limit, the su(2) algebra (\ref{su2}) written in terms of
$S^\pm$ and $S_3 \equiv \sigma_3$ contracts to the (projective)
e(2) (or Heisenberg-Weyl) algebra \cite{Wigner,Vitiello}
\begin{equation} \label{e2}
[ S_3,S^\pm] = \pm S^\pm \, , ~~~ [S^-, S^+] = 1 \, .
\end{equation}
The meaning of Eqs.~(\ref{relc}) and (\ref{e2}) is that, for large
$N$,~the laser excites collective dipole waves, $S^\pm$ denoting
the creation and annihilation operators of the associated quanta,
and the collection of single two-level particles manifests itself
as a bosonic system.

\section{Collective cooling of common vibrational modes} \label{sec3}

Each particle may couple to {\em its own} phonon modes and there
can also be {\em common} vibrational modes. We first discuss a
scheme for the collective cooling of common modes. This requires
the application of laser fields, each red-detuned by a phonon
frequency $\nu$ and with Rabi frequency $\Omega_\nu$. In the following, $b_\nu$ is the annihilation
operator for a phonon with $\nu$ and $c$ denotes the annihilation
operator for a cavity photon. The Hamiltonian of the system in the
interaction picture and within the rotating wave
approximation\footnote{This approximation introduces some errors
when calculating the behavior of the system for times $t \ll 1/
\nu$, with a restriction of achievable cooling rates from above.
Nevertheless, collective cooling can be much more efficient than
previously considered mechanisms \cite{new}.} then
equals\footnote{Note  that the coupling constant $g$ depends on
the geometry of the respective setup. In case of a ring cavity,
$g$ and the annihilation operator $c$ incorporate all possible
modes of the quantised electromagnetic field in the resonator.}$^,$\footnote{We also
observe that the cavity is in resonance with the atomic transition
of Figure 1(a) and therefore the cavity does not couple to the
vibrational modes of the particles.}
\begin{equation} \label{HI0}
H_{\rm I} = \sum_\nu {\textstyle {1 \over 2}} \hbar \sqrt{N}
\eta \Omega_\nu \, S^+ b_\nu + \hbar \sqrt{N} g  \, S^+ c +  {\rm H.c.}
\end{equation}
A detailed derivation of the atom-phonon coupling Hamiltonian in Eq.~(\ref{HI0}) can be found in Ref.~\cite{Gerry}. It applies in the Lamb-Dicke limit, where the atom-phonon coupling is relatively small compared to the phonon
frequency $\nu$ and $({\textstyle {1 \over 2}} \eta \Omega_\nu)^2 \ll \nu^2$. Moreover, we neglect the non-resonant coupling of the laser Hamiltonian to the 1-2 transition of the particles. This term is negligible compared to the driving of the resonant excitation of the sideband with coupling strength $\sqrt{N} {1 \over 2} \eta \Omega_\nu$ if $\Omega_\nu \ll \nu$. We neglect this non-resonant laser driving here since we do not expect it to have an effect on the conclusions drawn in the paper. It can only lead to an additional evolution between the states $|l \rangle_{\rm p}$ but cannot cause unwanted population outside the Dicke-symmetric subspace. Using the notation
\begin{eqnarray}
x_\nu \equiv \sqrt{N} \, {\textstyle {1 \over 2}} \eta  \Omega_\nu \, , ~~
x \equiv (\sum_\nu x_\nu^2)^{1/2} = {\textstyle {1 \over2}} \sqrt{N} \eta \Omega \, , ~~
b \equiv \sum_\nu (x_\nu/x) \, b_\nu \, , ~~
y \equiv \sqrt{N} g \, ,
\end{eqnarray}
Eq.~(\ref{HI0}) becomes
\begin{equation} \label{HI}
H_{\rm I} = \hbar x \, S^+ b + \hbar y \, S^+ c + {\rm H.c.} \, ,
\end{equation}
where the phonon annihilation operator $b$ obeys the familiar
commutator relation $[b,b^\dagger]=1$.

The leakage of photons through the cavity mirrors is accounted
for by considering the master equation \cite{textbook}
\begin{eqnarray} \label{master}
\dot \rho &=& - {{\rm i}\over \hbar} \, \left[ H_{\rm I}, \rho \right]
+ \kappa \, \big( c \rho c^\dagger - {\textstyle {1 \over 2}} \, c^\dagger c
\rho  - {\textstyle {1 \over 2}} \, \rho c^\dagger c  \big) \, ,
\end{eqnarray}
where $\rho$ is the density matrix of the combined state of all
particles,   the common vibrational mode and the cavity field.
Assuming regime (\ref{reg}) and restricting ourselves to
Dicke states with $l \ll N$, spontaneous emission from the
particles is negligible. Moreover, all noise terms (like heating)
with amplitudes small compared to $x$, $y$ and $\kappa$ can be neglected.

The concrete form of Eq. (\ref{master}) suggests  that the
stationary state $\rho_{\rm ss}$ corresponds to a coherent
state of the form $|\alpha \rangle_{\rm p} |\beta \rangle_{\rm v}
|\gamma \rangle_{\rm c}$ with $S^- |\alpha \rangle_{\rm p} =
\alpha \, |\alpha \rangle _{\rm p}$,  $b \, |\beta \rangle_{\rm v}
= \beta \, |\beta \rangle_{\rm v}$ and $c \, |\gamma \rangle_{\rm
c} = \gamma \, |\gamma \rangle_{\rm c}$. Indeed, the only solution
of $\dot \rho_{\rm ss} = 0$ is the state $|0 \rangle_{\rm p} |0
\rangle_{\rm v} |0 \rangle_{\rm c}$ with all particles in the
ground state, no photons in the cavity and no phonons in the
common vibrational mode. Once the atoms have been initialised, the
system loses its phonons within the time it takes to reach the
stationary state. Since this time evolution is solely governed by
the frequencies $x={1 \over2} \sqrt{N} \eta \Omega$, $y=\sqrt{N}
g$ and $\kappa$, we expect that this happens in the large $N$
limit in a time  given by the smallest of these frequencies.

To calculate the cooling rate  explicitly, we derive a set  of
differential equations for the variables
\begin{eqnarray} \label{coop1}
m \equiv \langle b^\dagger b \rangle_\rho \, , ~~
n \equiv \langle c^\dagger c \rangle_\rho \, , ~~
s_3 \equiv \langle S_3 \rangle_\rho \, , ~~~~~
\end{eqnarray}
where $m$ is the mean number of phonons with respect to the phonon
number operator $b^\dagger b$, $n$ is the mean number of photons
in the cavity mode and $s_3$ relates to the mean number of
particles in the excited state $|1 \rangle$. It is also useful to
consider the coherence quantities
\begin{eqnarray} \label{coop2}
k_1 &\equiv& \langle S^+b -S^- b^\dagger \rangle_\rho  \, ,~ \nonumber \\
k_2 &\equiv & \langle S^+c -S^- c^\dagger \rangle_\rho \, ,~ \nonumber \\
k_3 &\equiv & \langle b c^\dagger + b^\dagger c \rangle_\rho \, .
\end{eqnarray}
The introduction of (\ref{coop2}) is motivated by the fact that
the system first builds up the coherences $k_i$, which, once
established, provide an effective coupling between the different
subsystems. Using Eq. (\ref{master}) we obtain
\begin{equation} \label{dgl1}
\dot m = {\rm i} x k_1 \, , ~~
\dot n = {\rm i}y k_2 - \kappa n \, , ~~
\dot s_3 = - {\rm i} \, ( x k_1 + y k_2 ) \, .
\end{equation}
The annihilation and creation of phonons and photons is
accompanied by changes of the excitation number of the single
particles. For $l \ll N$, these fluctuations remains negligible on
average and the approximations $\langle S_3 b^\dagger b
\rangle_\rho  = s_3 m$,  $\langle S_3 c^\dagger c \rangle_\rho =
s_3 n$ and $\langle S_3 (b c^\dagger+b^\dagger c) \rangle_\rho  =
s_3 k_3$ can be adopted. We also neglect contributions of order
one compared to $N$ such as $\langle S^+S^- \rangle_\rho$. Then
\begin{eqnarray} \label{dgl2}
\dot k_1 &=& - {\textstyle {2{\rm i} \over N}}
\, ( 2 x m + y k_3 ) s_3 \, , \nonumber \\
\dot k_2 &=& - {\textstyle {2{\rm i} \over N}}
\, ( 2 y n + x k_3 ) s_3 - {\textstyle {1 \over 2}} \kappa k_2 \, ,
\nonumber \\
\dot k_3 &=& {\rm i} \, ( y k_1 + x k_2 ) - {\textstyle
{1 \over 2}} \kappa k_3 \, .
\end{eqnarray}
These non-linear differential equations imply
\begin{equation}
\label{sol} \dot m = {x \over 2y} \, \big( \kappa k_3 + 2 \dot k_3
\big) - {x^2 \over y^2} \, (\kappa n + \dot n) \, .
\end{equation}
We see below, that the presence of a negative $k_3$ and a positive
$n$ provides a cooling channel and plays a crucial role in the
cooling process.

\begin{figure}
\begin{minipage}{\columnwidth}
\begin{center}
\resizebox{\columnwidth}{!}{\rotatebox{0}{\includegraphics{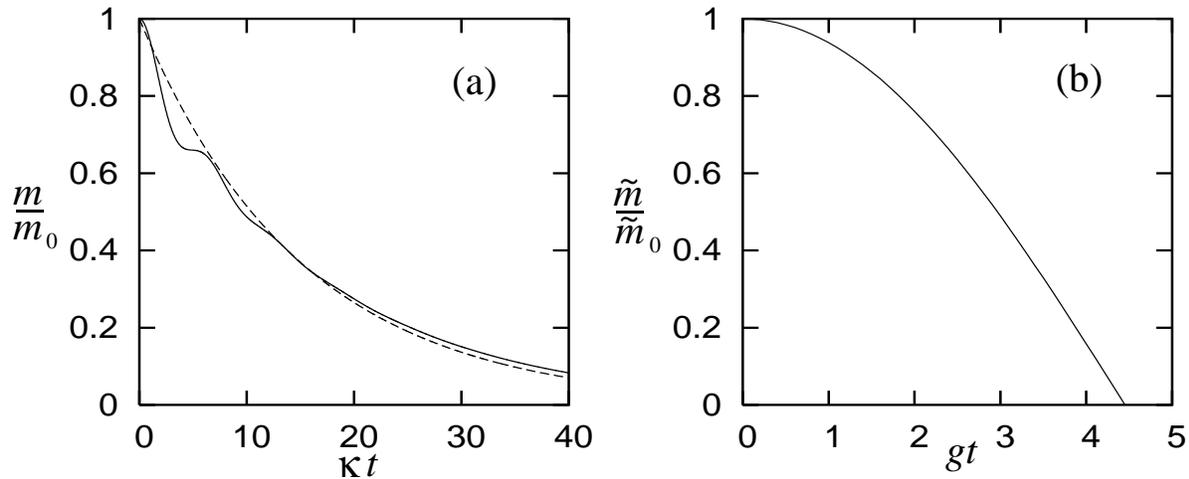}}}
\end{center}
\vspace*{-0.5cm} \caption{Cooling evolution of common vibrational
modes obtained from a numerical solution of
Eqs.~(\ref{dgl1})-(\ref{dgl2}) (solid line) in comparison to
Eq.~(\ref{m}) (dashed line) for $g=10^{-3} \, \kappa$,
$\eta\Omega=5 \cdot 10^{-4} \, \kappa$, $N=10^6$ and
$m_0=10^3$ (a). Cooling of individual phonon modes obtained from a
numerical solution of Eq.~(\ref{dgl1_indi}) assuming $\tilde n =
\tilde k_2 = 0$ and $\tilde s_3 =-{1 \over 2}N$ for $g=10^{-3} \,
\kappa$, $\eta\Omega=5 \cdot 10^{-4} \, \kappa$, $N=10^6$ and
$\tilde m_0 = 10^9$ (b).} \label{num}
\end{minipage}
\end{figure}

Here we are interested in the cooling of a large number of
particles. This allows us to solve the time evolution considering
first the regime where $\kappa \approx 0$ and Eq.~(\ref{sol})
becomes the conservation law
\begin{equation}
\dot m - {x \over y} \, \dot k_3 +  {x^2 \over y^2} \, \dot n =0 \, .
\end{equation}
In the parameter regime (\ref{reg}) and given that $m$, $n$  and
$s_3$ are of order $N$, the system reaches a stationary state with
$m$, $n$ and $k_3$ constant on a time scale of the order
$1/\sqrt{N}$. After such a time we might safely assume $\dot m =
\dot n = \dot k_1 = \dot k_2= \dot k_3 =0$, and obtain from
Eq.~(\ref{dgl2}) and for $\kappa \approx 0$ that $k_1= k_2 = 0$,
$k_3 = - (2 x/y) \, m$ and $n = (x^2/y^2) \, m$. This stationary
state is actually reached in a time given by the smallest among
$\sqrt{N}g$ and ${1 \over 2} \sqrt{N} \eta \Omega$, while
$\kappa \approx 0$ controls longer-lived processes such as the one
described by Eq.~(\ref{sol}). Under the above assumption for the
zeroth order in $\kappa$, we find $\dot m = - [x^2 (x^2 + y^2)/y^4]
 \, \kappa m$ to the first order in $\kappa$. From this we get
\begin{equation} \label{m}
m(t) = m_0 \, \exp \Big( - {x^2 (x^2 + y^2) \over  y^4} \, \kappa t \Big) \, ,
\end{equation}
where $m_0$ is the initial number of phonons in the system with
respect to the above defined operator $b^\dagger b$. The
exponential decrease of the phonon population (see Figure
\ref{num}(a)) amounts to the overall system cooling with a rate of
the same order of magnitude as $\kappa$. The result (\ref{m}),
which, we stress, holds under the condition of large $N$, shows
that the cooling of the system does not depend on the specific
value of $N$, and thus, provided $N$ is large, it holds even in
the case that not all the particles are initially prepared in
their ground state, which helps the feasibility of the scheme.
We also remark that such a behavior becomes possible only because
of dissipation, namely the leakage of photons through the cavity
mirrors.

\section{Collective cooling of the individual motion of the atoms} \label{sec4}

The collective regime established above may also be obtained in
the case where each particle $i$ couples to its {\em own set} of
individual phonon modes $b_{\nu,i}$. As above, we assume that the
Rabi frequencies  $\Omega_\nu$ of the corresponding laser fields
with detuning $\nu$ are for all particles the same. Then the
Hamiltonian of the system equals in the interaction picture
\begin{equation} \label{HI20}
H_{\rm I} = \sum_{\nu,i} {\textstyle {1 \over 2}} \hbar \eta
\Omega_\nu \, \sigma_i^+ b_{\nu,i} + \hbar g \, \sigma_i^+ c + {\rm H.c.}
\end{equation}
With the notation
\begin{eqnarray}
x_\nu \equiv \sqrt{N} \, {\textstyle {1 \over 2}} \eta  \Omega_\nu \, , ~~
x \equiv (\sum_\nu x_\nu^2)^{1/2} \, , ~~
b_i \equiv \sum_\nu (x_\nu/x) \, b_{\nu,i} \, , ~~
y \equiv \sqrt{N} g \, ,
\end{eqnarray}
where the $b_i$ obey~the relation $[b_i,b_j^\dagger]=\delta_{ij}$,
Eq.~(\ref{HI20}) simplifies to
\begin{equation} \label{HI2}
H_{\rm I} = {\hbar \over \sqrt{N}} \sum_i x \, \sigma_i^+ b_i + y
\, \sigma_i^+ c + {\rm H.c.}
\end{equation}
Suppose that the system is initially prepared in a state with  all
particles in the ground state, the mean phonon number is the same
for all particles and of about the same size as $m_0$ considered
before and there are no photons in the cavity mode. Then the
operator $\sum_{i=1}^N \sigma_i^+ b_i$ has a similar effect on the
system state as the operator $S^+ b$ in the previous case
\cite{piombino}. The net result is shown to be again
collective cooling in the large $N$ limit.

Proceeding as before, we first calculate the stationary  state
$\rho_{\rm ss}$. Leakage of photons through the cavity mirrors is
accounted for by using Eq. (\ref{master}). As before, the form of
the Eqs.~(\ref{master}) and (\ref{HI2}) suggests that $\rho_{\rm
ss}$ is the state with all particles in the ground state and no
phonons and no photons in the cavity. Indeed, it obeys $\dot
\rho_{\rm ss}=0$.

To calculate the cooling rate explicitly, we consider the
expectation values
\begin{eqnarray} \label{indi2}
\tilde m \equiv N \, \langle b_1^\dagger b_1 \rangle_\rho \, , ~~
\tilde n \equiv \langle c^\dagger c \rangle_\rho \, , ~~
\tilde s_3 \equiv N \, \langle \sigma_{31} \rangle_\rho
\end{eqnarray}
and the coherences
\begin{eqnarray}
\tilde k_1 &\equiv & \sqrt{N} \, \langle \sigma_1^+ b_1 - \sigma_1^-
b_1^\dagger \rangle_\rho \, , ~ \nonumber \\
\tilde k_2 &\equiv & \sqrt{N} \, \langle \sigma_1^+ c - \sigma_1^- c^\dagger
\rangle_\rho  \, ,  \nonumber \\
\tilde k_3 &\equiv & N \, \langle b_1 c^\dagger + b_1^\dagger c \rangle_\rho \, ,
\end{eqnarray}
where we sum over all particles. Using Eq. (\ref{master}) with the
Hamiltonian (\ref{HI2}) and the same approximations as before, we
obtain
\begin{eqnarray} \label{dgl1_indi}
\dot {\tilde m} = {\rm i} x \tilde k_1 \, , ~~
\dot {\tilde n} = {\rm i} y \tilde k_2 - \kappa \tilde n \, , ~~
\dot {\tilde s}_3 = - {\rm i} \, ( x \tilde k_1 + y \tilde k_2 )
\end{eqnarray}
and
\begin{eqnarray} \label{dgl1_indi2}
\dot {\tilde k}_1 &=& - {\textstyle {2{\rm i} \over N^2}} \,
( 2 x \tilde m + y \tilde k_3 ) \tilde s_3 \, , \nonumber \\
\dot {\tilde k}_2 &=& - {\textstyle {2{\rm i} \over N^2}} \,
( 2 Ny \tilde n + x \tilde k_3 ) \tilde s_3 - {\textstyle {1 \over 2}}
\kappa \tilde k_2 \, , \nonumber \\
\dot {\tilde k}_3 &=& {\rm i} ( y \tilde k_1 + x \tilde k_2 ) -
{\textstyle {1 \over 2}} \kappa \tilde k_3
\end{eqnarray}
implying
\begin{equation} \label{sol_indi}
\dot {\tilde m} = {x \over 2y} \, \big( \kappa \tilde k_3 + 2
\dot {\tilde k}_3 \big)
-  {x^2 \over y^2} \, (\kappa \tilde n + \dot {\tilde n}) \, .
\end{equation}
The differential equations (\ref{dgl1_indi})  and
(\ref{dgl1_indi2}) reveal that,  for $\kappa \approx 0$,  the
system reaches a stationary state within a time of the order one
with $\tilde k_1 = \tilde k_2 = 0$ and $\tilde k_3 = - (2x/y) \,
\tilde m$, given $\tilde m$ is of order $N$, while the cavity
accumulates a small population of photons. However,  $\tilde n
=(x^2/Ny^2) \, \tilde m$  remains small and the coherence $\tilde
k_3$ provides the main decay channel for the phonons in the
system. From Eq.~(\ref{sol_indi}) we obtain for large $N$, in
analogy to Eq.~(\ref{m}),
\begin{equation}
\tilde m(t) = \tilde m_0 \, \exp \Big( - {x^2 \over y^2}
\, \kappa t \Big) \, ,
\end{equation}
where $\tilde m_0$ is the initial total phonon number. Thus,
similar to the result (\ref{m}), the rate for the cooling of
individual phonon modes is of the same order of magnitude as the
cavity decay rate $\kappa$, after a transition time given by the
smallest among $g$ and ${1\over 2} \eta \Omega$ (see
Figure \ref{num}(b)).

\section{Conclusions}

In this paper we propose a new cooling mechanism based on the
collective excitation and de-excitation of particles trapped
inside an optical cavity. In Section \ref{sec2}, we showed that a
large atomic sample in the presence of highy symmetric
interactions, that treat all particles in the same way, behaves
like a collection of bosonic particles. In Sections \ref{sec3} and
\ref{sec4}, we then analysed the two extreme cases, where the
applied laser fields aim either at the cooling of only common or
individual motions of the particles. For both cases, we predict
cooling rates of the same order of magnitude as the cavity decay
rate $\kappa$. In the general case, one might argue that the
number of vibrational modes is extremely large. However, we
believe that similar cooling rates would be achievable in this
case as well, as long as the initial phonon energy in the setup
does not increase with the number of particles $N$ in the setup,
which is in general not the case.

We conclude with a few comments. The conservation law (\ref{sol})
obtained for $\kappa = 0$ allows to define the conserved quantity
$\dot Q =0$ with $Q \equiv m + (x^2/y^2) \, n - (x/y) \, k_3$. The
meaning of the time independence of  $Q$ is that one can shift the
quantities $k_3$, $m$ and $n$ by some constants without changing
the dynamics of the system apart from the changes in the process
of redistributing phonons, governed by $k_1$, and their population
$m$. The leakage of photons through the cavity mirrors ($\kappa
\neq 0$) disturbs the equilibrium expressed by $\dot Q=0$ inducing
a dynamical response, i.e.~a quantum phase transition: The overall
effect is the exponential decrease in the phonon population, namely
cooling \cite{piombino}.

The treatment in this paper should be compared with the
traditional approach where the attention is focused on the single
particle behavior. In such a single particle treatment
\cite{lewen} the cooling depends exponentially on the cavity decay
rate $\kappa$. In this case, $\kappa$ has to be about the same
size as $g$ and ${1 \over 2} \eta \Omega$, which imposes a
strong constraint on the time scale of the cooling process. The
advantage of our treatment lies, on the contrary, in the fact that
the numbers of particles $N$ introduces, in principle, the crucial
freedom to tune $\kappa$ to be as large as $\sqrt{N}g$ and  ${1
\over 2} \sqrt{N} \eta \Omega$ thus cooling the particles
very efficiently. Although corrections might be necessary in
realistic situations, a high cooling rate also reduces the number
of collisions between particles, which certainly are a problem in
less effective cooling schemes. We believe that considering the
cavity-mediated collective field theoretical behavior opens a new
perspective in particle cooling. \\[0.5cm]

{\em Acknowledgement}. This work was supported in part by the
European  Union, COSLAB (ESF Program), INFN, INFM  and the UK
Engineering and Physical Sciences Research Council.  A.B.
acknowledges stimulating discussion with Philippe Grangier and
thanks the Royal Society and the GCHQ for funding as a James Ellis
University Research Fellow.

\end{document}